# Superconducting Quantum Interference Device based on MgB$_2$ nanobridges

A. Brinkman, D. Veldhuis, D. Mijatovic, G. Rijnders,
D. H. A. Blank, H. Hilgenkamp and H. Rogalla.

*Low Temperature Division, Faculty of Applied Physics, and MESA$^+$ Research Institute,
University of Twente, P.O. Box 217, 7500 AE, Enschede, The Netherlands.*

**The recently discovered superconductor MgB$_2$, with a transition temperature of 39K, has significant potential for future electronics. An essential step is the achievement of Josephson circuits, of which the superconducting quantum interference device (SQUID) is the most important. Here, we report Josephson quantum interference in superconducting MgB$_2$ thin films. Modulation voltages of up to 30 μV are observed in an all-MgB$_2$ SQUID, based on focused ion beam patterned nanobridges. These bridges, with a length scale < 100 nm, have outstanding critical current densities of 7 x 10$^6$ A/cm$^2$ at 4.2 K.**

Following the surprising discovery by Nagamatsu and Akimitsu *et al.*[1] of superconductivity at a temperature of 39 K in magnesium-diboride, numerous groups have started to investigate MgB$_2$ due to its great potential for large current and electronic applications. The reasons for the interest in using MgB$_2$ in superconducting electronics are manifold. The large charge carrier density[2], the isotropy of the material[3] and the fact that grain boundaries in polycrystalline MgB$_2$ are strong links[4,5] are important advantages as compared to the cuprate high temperature superconductors. In comparison with the conventional metallic superconductors, MgB$_2$ has the potential for higher operation speed due to its larger energy gap[6]. Finally, the high transition temperature of MgB$_2$ facilitates cooling of superconducting electronic circuits by cryocoolers. Essential for the realization of superconducting electronics is the availability of high quality thin films and the technology to fabricate Josephson circuits in these films. Several groups have succeeded recently in fabricating superconducting MgB$_2$ films[7-13]. The procedures used for this are not foreseen to be suited for the realization of trilayer Josephson junctions because of the post-anneal step involved. Fortunately, aside from Josephson junctions that include a barrier-layer, also nanobridges can be employed as the weak links in a SQUID[14]. In the following, we present the observation of Josephson quantum interference in MgB$_2$ using two superconducting nanobridges of 150 nm by 70 nm incorporated in a superconducting ring.

In Fig. 1, a Scanning Electron Microscopy image of a nanobridge in a MgB$_2$ film is depicted. When a transport-current is directed through the bridge a magnetic field is created, which can penetrate the superconductor in the form of Abrikosov vortices if the field is larger than the lower critical field ($H_{c1}$). Two of such vortices, with opposite orientation, will then be created simultaneously at the edges of the bridge[15]. An edge-pinning force will act on the vortices, but with increasing transport current the Lorentz force acting on the vortex will overcome this pinning. Consequently, for bias currents exceeding the critical current $I_c$, the vortices will move towards each other and finally annihilate. In the process of vortex motion an electric field is induced in the direction of the transport current and energy dissipation will take place.

The Abrikosov vortices have a normal core of radius $\xi$, the coherence length, which for $MgB_2$ is reported to be 5.2 nm[5]. For superconducting bridges that are smaller than a few times $\xi$, the normal core area in the bridge will act as a Josephson weak link, with a characteristic relationship between the supercurrent and the phase change of the macroscopic wave function over the weak link[16]. Importantly, also wider bridges can show a significant current-phase relationship, provided the width of the bridge is comparable to, or smaller than, the effective London penetration depth $\lambda_\perp$[17]. Reported values for the bulk penetration depth for $MgB_2$ vary from 140 to 180 nm[3, 5], yielding an effective penetration depth, $\lambda_\perp = \lambda_L \coth(d/2\lambda_L)$ of 230 to 360 nm for films with a thickness $d$ of 200 nm.

In a superconducting ring, the total phase-change of the superconducting wave function when going around the hole is quantized in multiples of $2\pi$. With the nanobridges incorporated in the ring, the phase-change is composed of two contributions. The first is due to the current flow through the nanobridges; $\Delta\varphi_1(I_1)$ and $\Delta\varphi_2(I_2)$, with $I_1$ and $I_2$ the currents through bridges 1 and 2, respectively, and the second is associated with the applied magnetic flux $\Phi$ in the ring. With this, the quantization-condition becomes $\Delta\varphi_1 - \Delta\varphi_2 + 2\pi(\Phi/\Phi_0) = 2\pi k$, with k an integral number and $\Phi_0$ the elemental flux quantum ( = $2.07 \times 10^{-15}$ Tm$^2$). The ring is superconducting for bias currents $I_{bias} = I_1 + I_2$, for which this condition can be fulfilled. By varying the applied flux, the maximal attainable $I_{bias}$ to fulfill the quantization-condition is modulated with a period $\Phi_0$. This critical current will be maximal when the enclosed flux equals a integral number n times $\Phi_0$ and is minimal for $\Phi = (n+1/2)\Phi_0$. This is the basic principle underlying the operation of a dc SQUID.

For the fabricaton of the SQUIDs, first, thin films of 200 nm $MgB_2$ were deposited on MgO substrates, by pulsed laser ablation in a two-step in-situ process[11]. The $MgB_2$ is deposited at room temperature from an Mg-enriched $MgB_2$ target and subsequently annealed at 600 $^o$C. The transition temperature $T_c$ of the as-deposited films is 24 K[18]. These films are polycrystalline. This does not hamper the supercurrent, since the grain boundaries in $MgB_2$ act as strong links[4, 5].

The SQUID and contact-paths were patterned in two steps. First, the coarse structures, including the square-washer SQUID-ring and the contact leads were defined by standard photolithography and argon ion-beam milling. With an acceleration voltage of 500 V, the ion milling under an angle of 45° occurs at an etching rate of approximately 5 Å/s. The SQUID, with an estimated inductance of 60 pH, consists of a square-washer of 20 x 20 μm inner and 70 x 70 μm outer dimension and a 5 x 55 μm slit, as is shown in Fig. 2(a). Further, the structure contains two striplines, 30 μm long and 5 μm wide, into which, subsequently, nanobridges were structured by direct FIB milling. Using a 25 kV Ga$^+$ beam with a diameter of 50 nm (Full Width at Half Maximum) and a beam current of 40 pA, trenches are etched in the MgB$_2$ films at a rate of 0.30 mm$^3$/Coulomb. The density of the Ga$^+$ ions has a Gaussian distribution in the central part of the beam. The nanobridges are made by letting two beam profiles partly overlap, which results in a reduced height of the bridge, as compared to the original film thickness. The dimensions of the fabricated nanobridges are shown in Fig. 2(b); the width is about 70 nm (FWHM) and the height of the nanobridges is approximately 150 nm. The length of the bridges is 150 nm.

The electrical transport properties of the ring-structure were measured in a four-point configuration in a shielded variable-temperature flow-cryostat. The transition temperature of the structure was found to be 22 K, comparable to the original $T_c$ value of the unpatterned film. In Fig. 3(a), a typical example of the measured SQUID current-voltage characteristics is shown for the two extremal values of the enclosed magnetic flux of the SQUID, at a temperature $T = 19$ K. Above $T = 12$ K the current-voltage characteristics are non-hysteretic, with a parabolic shape of the voltage branch. Below 12 K a hysteresis appears, as can be seen in the inset of Fig. 3(a) where a current-voltage characteristic at 10 K is shown. This hysteresis is presumed to arise from the considerable heating of the bridges by the large bias-currents needed at these temperatures. In Fig. 3(b), the measured critical current dependence on temperature is depicted. For a rounded bridge edge, as is the case here, the critical current is expected to be proportional to $\lambda^{-2}$ and $\xi^{-1}$, as described in Ref. 15. Given the temperature dependencies for $\lambda$ and $\xi$ from the two-fluid model in the clean limit[19], $\lambda = \lambda_0(1-t^4)^{-1/2}$ and $\xi = \xi_0(1-t)^{-1/2}$, where $t = T/T_c$, the critical current is expected to behave as $I_c = I_{c0}(1-t^4)^{1/2}(1-t)^{1/2}$, which fits the observed dependence very well, as is shown in Fig. 3(b).

The critical current of the SQUID at $T = 4.2$ K is 1.5 mA. With an estimated bridge cross-section of 70 x 150 nm this results in a critical current density of 7 x 10$^6$ A/cm$^2$. This large value implies that the nanostructuring by the use of a Ga focused ion beam is very well possible while maintaining large critical current densities. This is important information, since it shows that the chemical reactivity and volatilaty of the magnesium do not pose problems in the nanostructuring. Furthermore, we note that the structures are very stable over time and are insensitive to thermal cycling or exposure to moisture.

In Fig. 4, the voltage modulation of the SQUID at different constant bias currents is shown as function of the applied magnetic field. The period of the modulation in Fig. 4 is 0.74 µT. With a period of $\Phi_0$ for the SQUID critical current modulation by applied magnetic flux, an effective SQUID area, $A_{eff} = \Phi_0/H$, of 2.8 x 10$^3$ µm$^2$ is obtained, which is in well accordance with the actual SQUID dimensions, taking flux-focusing by the superconducting washer into account[20]. Above the temperature at which the current-voltage characteristics become hysteretic, the voltage modulation shows the same temperature dependence as the critical current. A modulation voltage of 30 µV was observed at 10 K. Voltage modulation was observed up to 20 K.

The presented results prove that high-quality superconducting structures can be realized on a lengthscale < 100 nm in in-situ fabricated MgB$_2$ films. MgB$_2$ ring-structures incorporating nanobridges display Josephson quantum interference effects, which forms the basis for the creation of an all MgB$_2$ SQUID. This result is an essential step towards sensors and electronic circuits based on this novel superconductor.

The authors thank H. J. H. Smilde, M. Yu. Kupriyanov, A. A. Golubov, and G. J. Gerritsma for useful discussions. This work was supported by the Dutch Foundation for Research on Matter (FOM). H. H. acknowledges support by the Royal Dutch Academy of Arts and Sciences.

**Figure captions**

**FIG. 1**. Scanning Electron Microscopy image of a $MgB_2$ nanobridge. The width of the bridge is approximately 70 nm, the length 150 nm.

**FIG. 2.** (**a**) Schematic lay-out of the SQUIDs and (**b**) schematic cross-section of the nanobridges.

**FIG. 3**. (**a**) Current-voltage characteristics of a SQUID at $T = 19$ K for different values of the enclosed magnetic flux. In the inset, a hysteretic current-voltage characteristic is shown for $T = 10$ K; the kinks in the voltage branch indicate the onset of additional channels for vortex flow[17,21]. (**b**) Critical current of the SQUID as function of temperature. The dashed line shows the theoretical fit.

**FIG 4**. SQUID voltage modulation at 15 K, at different values of the current bias.

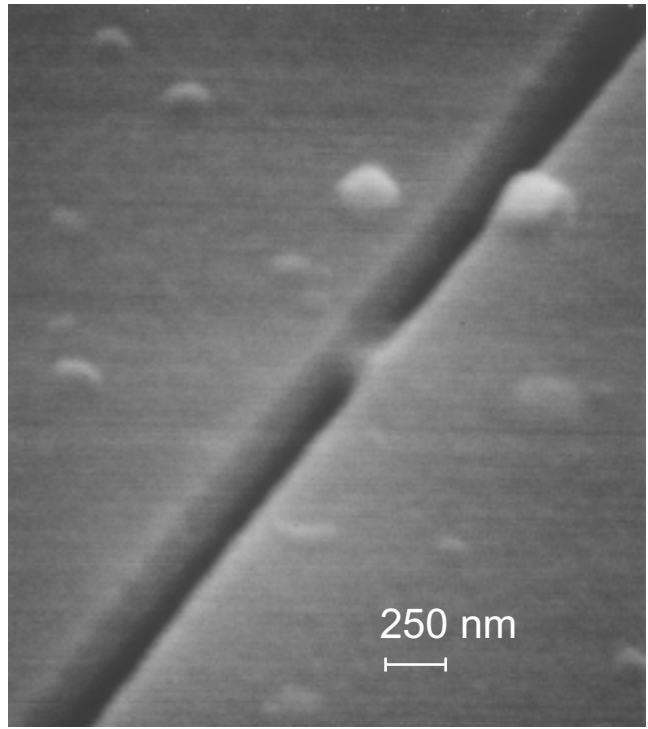

Figure 1

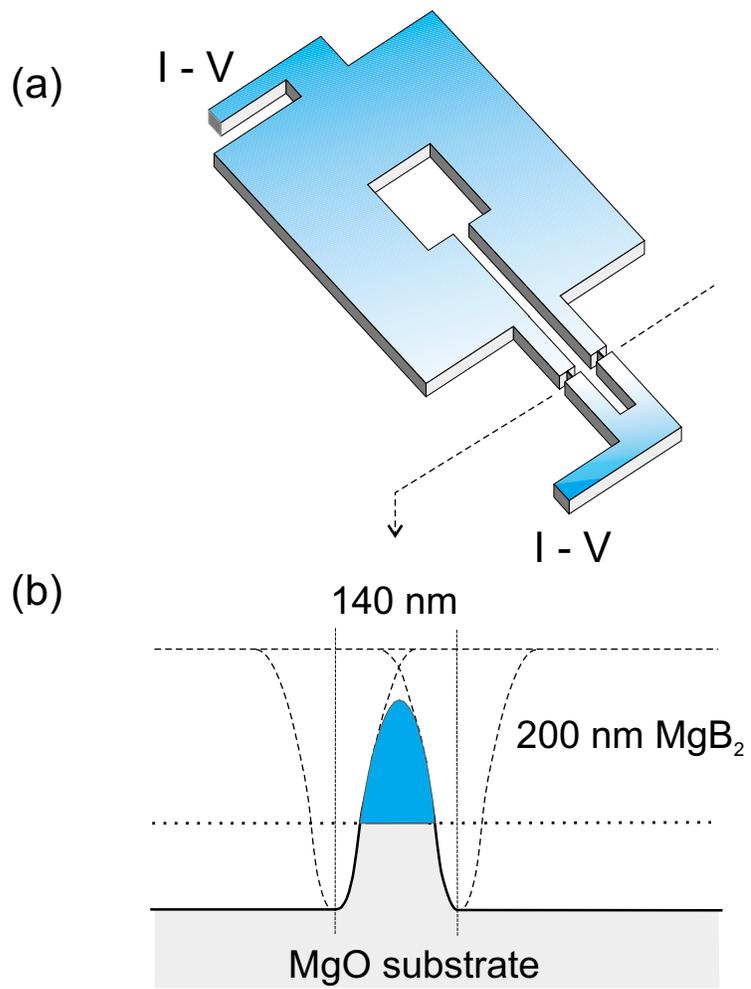

Figure 2

(a)

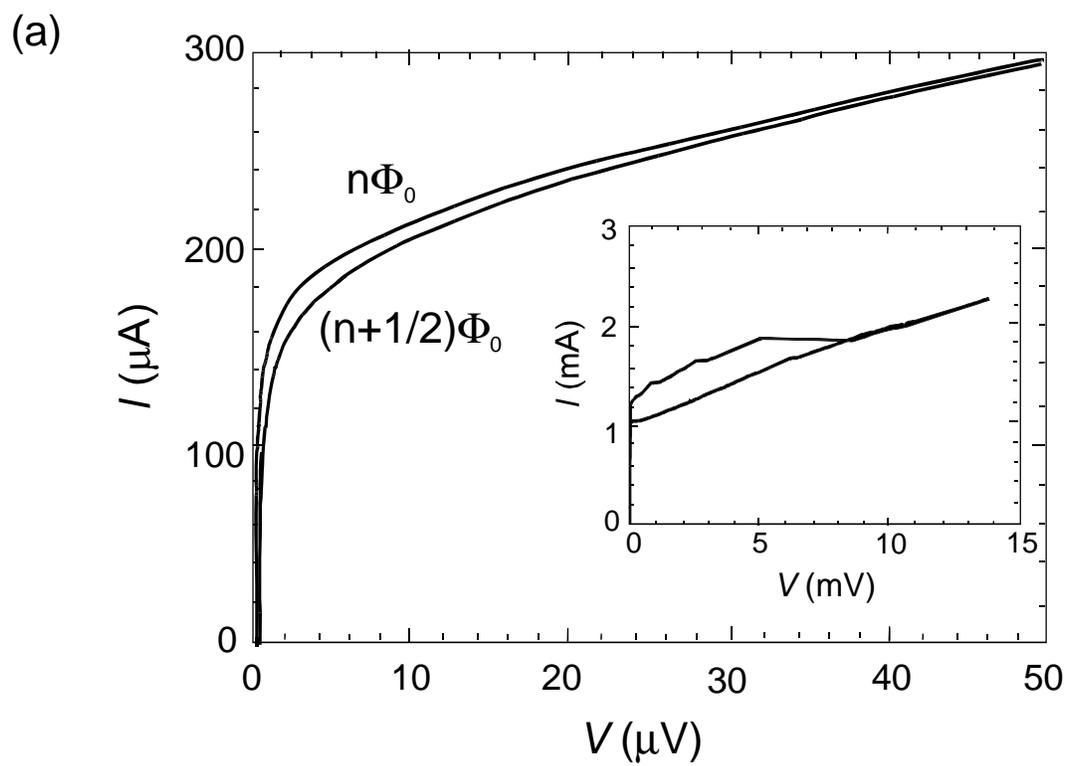

(b)

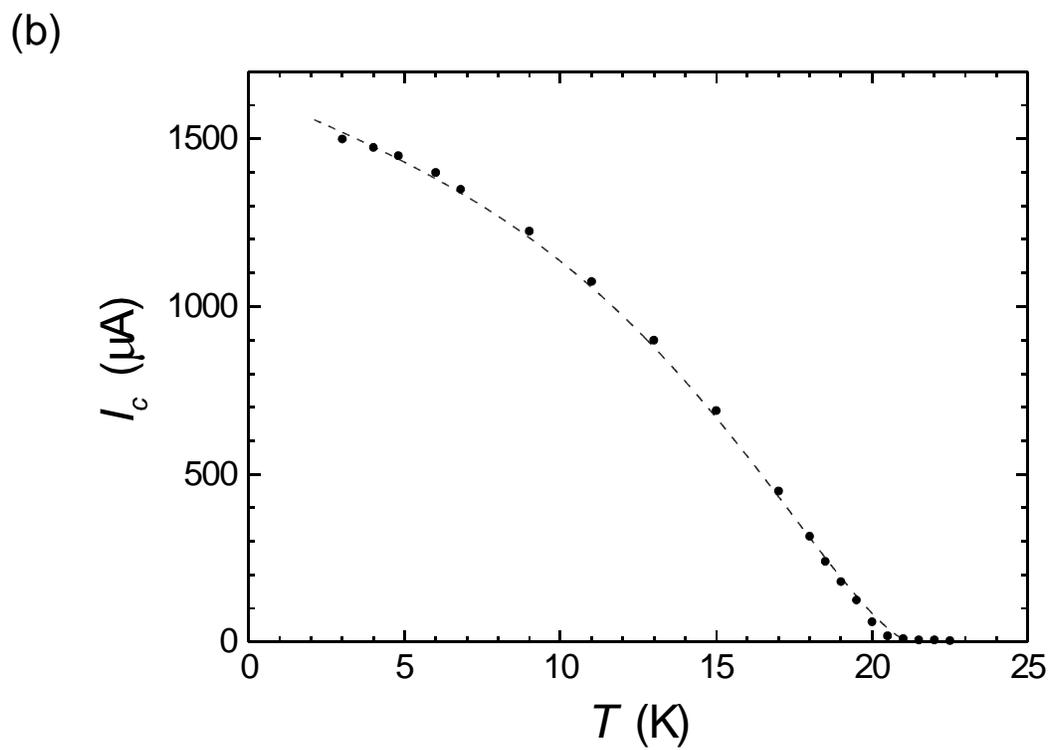

Figure 3

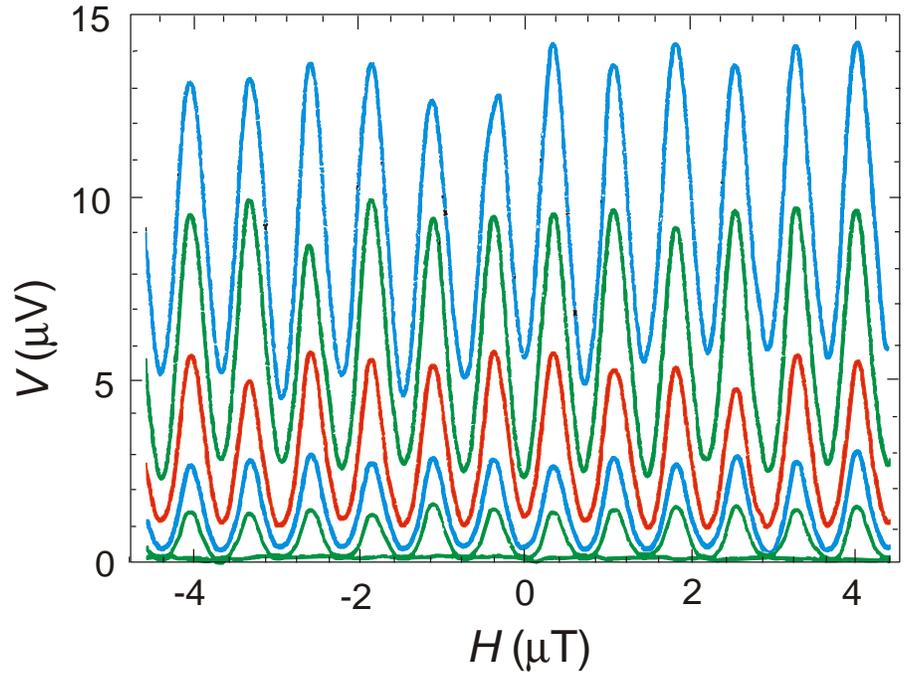

Figure 4